\documentclass[journal,twoside]{IEEEtran}
\usepackage{cite}

\ifCLASSINFOpdf
   \usepackage[pdftex]{graphicx}
   \graphicspath{{./PDF/}{./JPG/}{./PNG/}{./TIF}}
   \DeclareGraphicsExtensions{.pdf,.jpg,.png,.tif}
\else
\fi

\usepackage{amsmath}
\usepackage{array}
\newcolumntype{C}[1]{>{\centering\let\newline\\\arraybackslash\hspace{0pt}}m{#1}}
\usepackage[export]{adjustbox}
\usepackage[caption=false,font=footnotesize]{subfig}
\usepackage{float}
\usepackage{dblfloatfix}

\usepackage{multirow}
\usepackage{gensymb} 
\usepackage{url}     
\usepackage[usenames, dvipsnames]{color}

\usepackage{lipsum}
\usepackage{textcomp}
\usepackage[capitalise,nameinlink]{cleveref}

\usepackage{booktabs, cellspace, hhline}
\usepackage{bm}
\usepackage{xfrac}
\def\BibTeX{{\rm B\kern-.05em{\sc i\kern-.025em b}\kern-.08em
    T\kern-.1667em\lower.7ex\hbox{E}\kern-.125emX}}
    
\begin{document}

\title{{\fontsize{24}{26}\selectfont{Communication\rule{29.9pc}{0.5pt}}}\break\fontsize{16}{18}\selectfont
Multi-Beam Graded Dielectric Lens Antenna from Multi-Material 3D Printing}

\author{Henry~Giddens
	and~Yang~Hao
	
	\thanks{Manuscript received XXXX; revised XXXX.}
	\thanks{This work was funded under the Adaptive Communications Transmission Interface (ACTI) Olympia project by the Defence Science and Technology Laboratory (DSTL). The authors would like to thank Marcus Walden, Thomas Rouse and Simon Bates from Plextek for their input and support.}
	\thanks{H. Giddens (h.giddens@qmul.ac.uk) and Y. Hao (y.hao@qmul.ac.uk) are with the School of Electronic Engineering and
		Computer Science at Queen Mary University London.}%
	\thanks{}
}

\markboth{}
{Multi-Beam Graded Dielectric Lens Antenna from Multi-Material 3D Printing}

\maketitle

\begin{abstract}
Modern communication systems will require antennas with adaptive functionality which are able to modify their performance based on the requirements of the channel. For example, mobile ad-hoc networks need directive antennas that are able to radiate in any direction across the 360$\degree$ azimuth plane. Conformal antennas that can be simply operated to have multi-functional performance characteristics are therefore of interest. In this paper, we present a gradient-index lens antenna designed to radiate with a 45$\degree$ beamwidth across 8 different sectors. When fed by a simple switched feeding network, the lens is able to provide 360$\degree$ azimuth coverage in 45$\degree$ segments. Further analysis of the radiation patterns shows how two distinct multi-beam patterns can be produced from simple feed networks and simultaneous excitation of each feeding element. The proposed lens is fabricated by multi-material 3D printing. The final lens radiates with a gain of 8.5 dBi when a single sector is excited, and with a maximum gain of 5.9 dBi in multi-beam mode. Finally, it is shown how the lens can also radiate omnidirectionally when optimised phase and amplitude weightings are applied to each port.
\end{abstract}


\section{Introduction}\label{sec1}

\IEEEPARstart{R}{ecent} advances in modern communications systems, such as the rapid development of software defined radios, will require antennas with multi-functionality that are able to operate in ever-changing wireless channels \cite{1336725,tawk2016antenna}. For example, next generation mobile ad hoc network (MANET) radios for military use will require directional antennas to overcome current performance limitations which restrict the range of the radio link and are significantly affected by self-interference \cite{923548,4799780,10.1007/978-3-540-85209-4_5}. Such antennas, when placed on mobile terminals, require radiation pattern diversity as they must be able to communicate with a terminal in any direction at a single point in time. The ability to dynamically modify the radiation pattern of an antenna positioned on the mobile terminal would enable communication with multiple users whilst mitigating interference effects from unwanted users or multi-path scatterers. In such systems there is a requirement for antennas with the ability to dynamically adapt their radiation patterns with respect to the instantaneous requirements of the communication system. Recent research has focussed on new antenna architectures with 360$\degree$ steerable and switchable beams \cite{2617820,8052527,8219414,8542736,8396302}, which complement traditional techniques such as electronically steerable passive array radiator (ESPAR) antennas \cite{1487777}, switched and full phased arrays \cite{1512075,6069837}.

Lens antennas are a particularly useful approach for generating optimised radiation patterns. Lenses with multiple focal points and beam steering ability have long been of interest \cite{1954JAP_Gutman,1143391}, whilst in recent years, a number of novel lenses  have been developed and realised practically \cite{doi:10.1002/admt.201600072,TO-YHAO,TAP:2013:2282905,6266701,PhysRevB.84.165111,6719535,Yang:11}, including multi-beam collimators which can be realised by near-zero index metamaterials \cite{TAP:2013:2282905,6266701,PhysRevB.84.165111}, and a flat aperture Luneburg lens with beam steering capability \cite{6719535}. Integrated lens antennas with directly coupled feeds provide an advantageous solution to feeding the lens, delivering high directivity, good aperture efficiency and low side lobe levels which are particularly important for highly directive beams at mm-wave frequencies, although the integrated feeding structure can limit the operational bandwidth \cite{4463919}. Antenna lenses can be designed from all-dielectric media via quasi-conformal coordinate transformations \cite{TO-YHAO}.

In order to achieve graded-index (GRIN) material profiles practically, ceramic composites with customisable dielectric constants have been developed and integrated within the profile of such GRIN devices \cite{6719535,Yang:11}. However this technique requires time and specialised resources, and is unsuitable for rapid prototyping. Using the effective-medium theory, GRIN materials have been realised by including sub-wavelength air-gap intrusions within a host material in order to lower the effective dielectric constant. The method has been demonstrated to work effectively in EM devices such as Luneburg lenses, where the dielectric profile is required to vary radially so that the lens can be fed from any direction \cite{10.1038/ncomms1126}.
More recently, 3D printing has emerged as an attractive additive manufacturing process for antennas designed at microwave and millimetre-wave frequencies as it allows engineers to design and fabricate antennas with finely detailed features \cite{10.1038/ncomms1126}. 3D printing also offers the possibility to achieve effective graded-index profiles, where a unit cell volume is only partially filled with the 3D filament, the rest of the volume occupied by air. In such a case, the overall permittivity of the unit cell is lower than that of the 3D filament and by increasing the filling factor of the air gap within the unit cell, 3-dimensional GRIN materials can easily be realised \cite{doi:10.1002/admt.201600072,10.1038/ncomms1126,LARIMORE201748}. Filament materials with high dielectric constants are now also available, making the realisation of new antennas possible \cite{101038,doi:10.1002/admt.201600072,PREPERM}. 

In this communication, a multi-functional 8-segment graded-dielectric lens antenna is presented. Using a distributed and isolated monopole feed network with one port per segment, a TO-inspired graded index lens is designed to shape the radiation from each segment to a 45$\degree$ half-power beamwidth. The lens is able to radiate with a distinct multi-beam radiation pattern when the ports are fed simultaneously with simple phase weighting. Furthermore, the lens can radiate omnidirectionally  with less than 3 dB ripple in the H-plane pattern when phase and amplitude to each port is controlled. The proposed application for such an antenna is a mobile terminal in a MANET radio which must isolate radiation to terminals in a particular direction, mitigating interference from other sources. In such a system, the antenna may also be required to radiate omnidirectionally in order to locate the relevant terminals.

\section{Design of 2-Dimensional 8-segment All-Dielectric Lens Antenna}\label{sec2}
\begin{figure}
    \centering
    \subfloat[]{%
	\includegraphics[width=0.45\linewidth,valign=b]{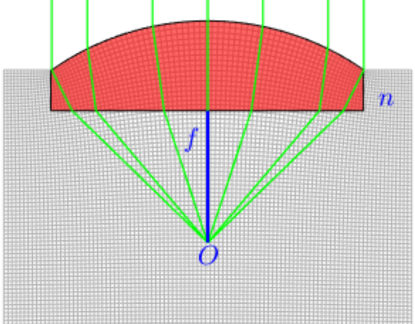}%
	\vphantom{\includegraphics[width=0.45\linewidth,valign=b]{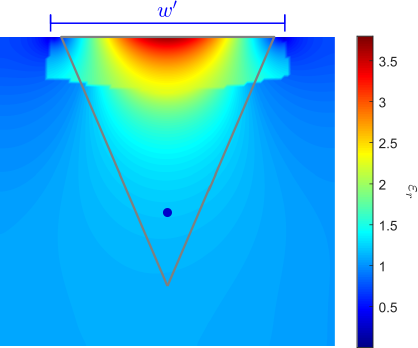}}%
	\label{Fig1a}%
	}%
	\hfill
	\subfloat[]{%
	\includegraphics[width=0.45\linewidth,valign=b]{Fig1-B.pdf}%
	\vphantom{\includegraphics[width=0.45\linewidth,valign=b]{Fig1-B.pdf}}%
	\label{Fig1b}%
	}%
	\\
	\subfloat[]{%
	\includegraphics[width=0.4\linewidth,valign=c]{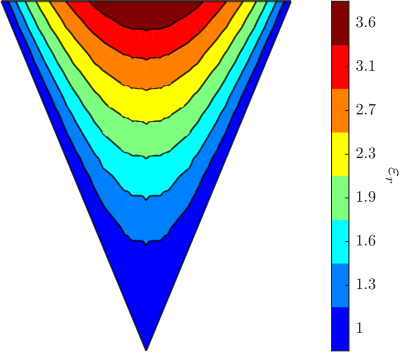}%
	\vphantom{\includegraphics[width=0.6\linewidth,valign=c]{Fig1-C.pdf}}%
	\label{Fig1c}%
	}%
	\hfill
	\subfloat[]{%
	\includegraphics[width=0.6\linewidth,valign=c]{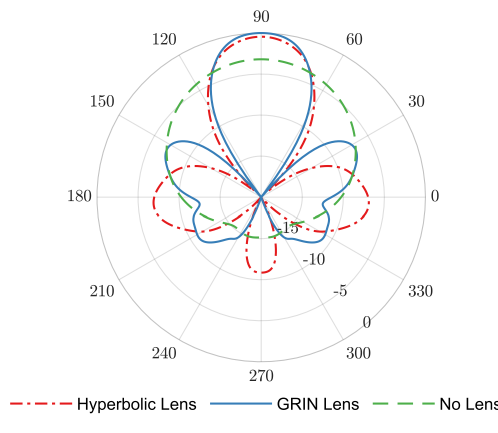}%
	\vphantom{\includegraphics[width=0.6\linewidth,valign=c]{Fig1-C.pdf}}%
	\label{Fig1D}%
	}%
	\caption{(a) Hyperbolic lens. (b) Permittivity map of transformed GRIN lens overlaid with outline of octagonal sector. (c) Discretized permittivity values of lens sector. (d) Radiation patterns from original hyperbolic lens, GRIN lens, and waveguide port on its own}
    \label{Fig1}
\end{figure}

The basic function of an antenna lens is to alter the propagation path length from a source located at the focal point (or feed position) so that the phase front of the electromagnetic wave emerging from the lens is aligned. Here, a GRIN lens was designed using a quasi-conformal coordinate transformation to convert a hyperbolic lens on to a flat surface. The entire media incorporating the lens was modified in order to lie within a single sector of an octagon. \cref{Fig1a} shows a diagram of the hyperbolic lens designed to produce a 45$\degree$ half-power beam width from a point source located at the focal-length, $f$. The initial hyperbolic lens had a relative permittivity of 4, radius of curvature of 100 mm and a width of $w$ = 75 mm. The focal point was at a position of $f$ = 33 mm. These parameters are important in ensuring the lens is correctly illuminated by the source and effectively determine the performance of the lens \cite{4907139}. The well-known quasi-conformal mapping technique was then used to modify the geometry of the hyperbolic lens, creating a flat aperture with a reduced overall footprint, resulting in a graded-dielectric material profile, which is shown in \cref{Fig1b}. The transformed space was expanded by a factor of 1.15 to ensure the background medium around the focal point had a relative permittivity close to 1. \cref{Fig1b} is overlaid with the outline of a sector representing $\sfrac{1}{8}$\textsuperscript{th} of an octagon, where the internal angle is equal to 45$\degree$. In order to minimize the effects arising from the areas where $\varepsilon_r$ \textless 1, the outline of this sector was set to overlay the lens by a slightly smaller width than the limits of the original lens. The permittivity map of \cref{Fig1b} was discretized into the different values displayed in \cref{Fig1c}. This was done by selecting different areas where the permittivity values differed by less than 12.5\% and then selecting the average value for that sector. All values less than 1.3 set equal to 1. At the edge of the lens, the maximum permittivity was 3.6 and at the centre of the lens this reduced to a value of 1. The transformed GRIN lens had an aperture width of $w'$ = 86 mm, which was reduced to 70 mm after discretization. The length of the discretized lens was 84 mm.

\section{Radiation Pattern Analysis of an Octagonal Multi-Functional Lens Antenna}\label{sec3}
A parametric sweep was run to determine the dimensions of the hyperbolic lens required to radiate with a beamwidth of 45$\degree$ at the target frequency of 5.8 GHz. The initial simulations were performed in a quasi-2D scenario with electrical boundary conditions terminating the simulation space in the +/- $z$ directions, and PML boundaries in the $x$ and $y$ directions. A waveguide port with a width of 20 mm was placed at the focal point. \cref{Fig1D} shows the 2D far field radiation patterns of the hyperbolic and GRIN lenses and the waveguide port alone. As can be seen, a main beam with half-power beam width of 45$\degree$ is produced by both lenses. An improvement in the normalized gain of 3.2 dBi is achieved by both lenses when compared to the radiation from the port alone.

\begin{figure}[!t]
	\centering
	\hspace{0.25cm}
	\subfloat[]{%
		\includegraphics[width=0.4\linewidth,valign=c]{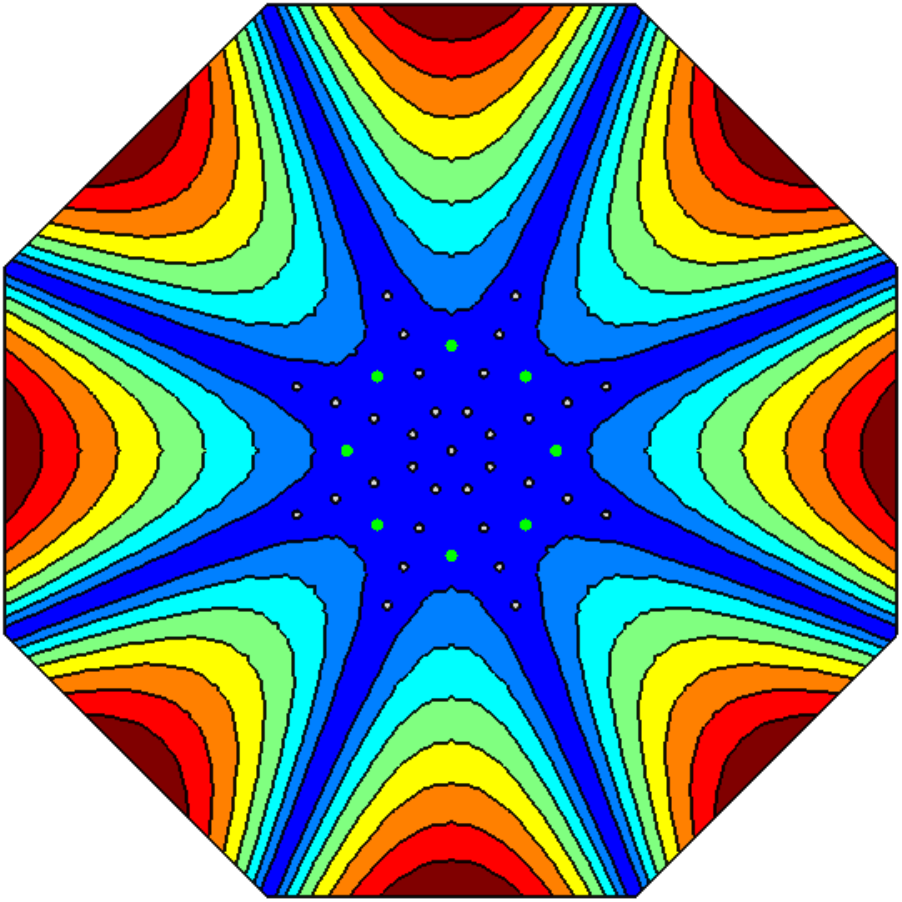}%
		\vphantom{\includegraphics[width=0.4\linewidth,valign=c]{Fig2-A.pdf}}%
		\label{Fig2a}%
	}%
	\hfill
	\subfloat[]{%
		\includegraphics[width=0.45\linewidth,valign=c]{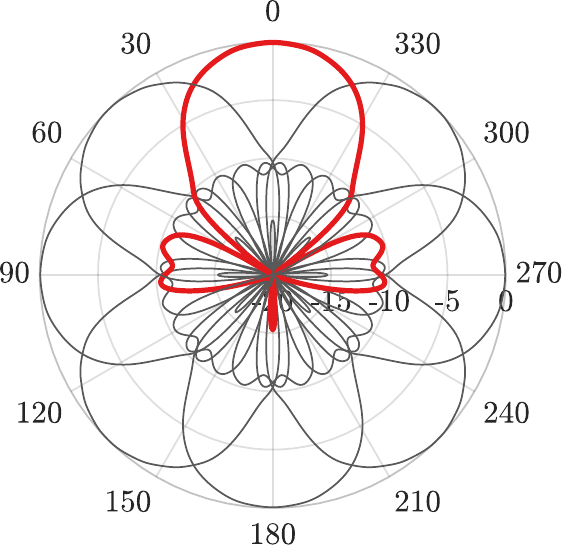}%
		\vphantom{\includegraphics[width=0.4\linewidth,valign=c]{Fig2-A.pdf}}%
		\label{Fig2b}%
	}%
	\hspace{0.25cm}
	\caption{(a) The full octagonal graded dielectric lens with 8 individual segments - the feed points are shown in green. (b) The 2D radiation patterns from each segment at 5.8 GHz.}
	\label{Fig2}

	\centering
	\hspace{0.5cm}
	\subfloat[]{%
		\includegraphics[height=3.5cm,valign=c]{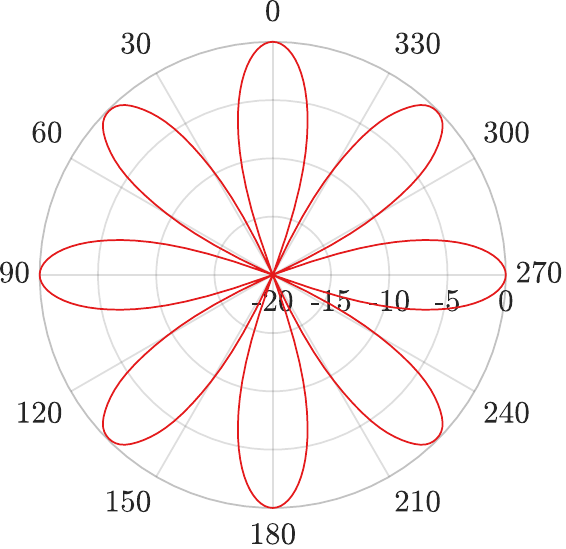}%
		\label{Fig3a}%
	}%
	\hfill
	\subfloat[]{%
		\includegraphics[height=3.5cm,valign=c]{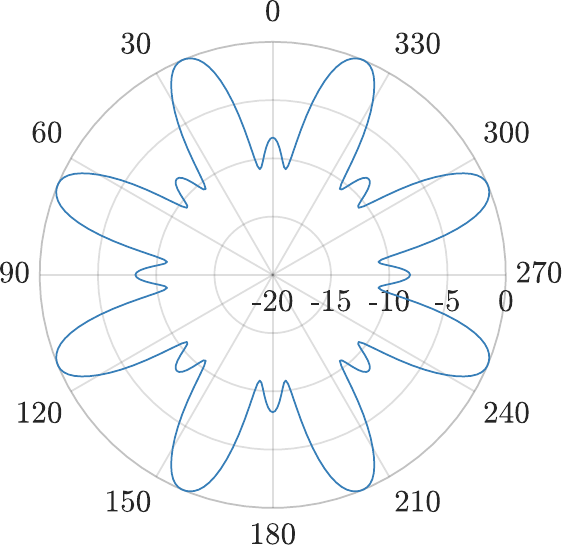}%
        \label{Fig3b}%
	}%
	\hspace{0.5cm}
	\caption{Radiation patterns of the 2D 8-sector lens with multi-port excitation. (a) Anti-phase excitation. (b) Equal-phase excitation.}
	\label{Fig3}%
\end{figure}

The GRIN lens segment shown in \cref{Fig1c} was rotated through 45$\degree$ and copied seven times, to form a full octagonal lens able to switch between different beams covering the entire 360$\degree$ azimuth plane. Each segment of the octagonal lens was excited with a separate feed. The waveguide ports from the previous simulations were replaced with a point source positioned at the focal point. Each segment was isolated from its neighbours by 4 reflectors with an inter-element spacing of 8 mm. The reflecting elements had a diameter of 3 mm. The full octagonal lens is shown in \cref{Fig2a}. The lens had $x$-$y$ dimensions of 172 x 172 mm and an aperture width of 70 mm for each segment.The radiation patterns of each segment are shown in \cref{Fig2b}. Each segment lens produced a pattern with a half-power beamwidth of 50$\degree$, with adjacent beams crossing over at the -1.5 dB level at an angle of 22.5$\degree$. The side lobe levels were always less than 10 dB below the main beam maximum with very little backward radiation due to the isolation provided by the reflector elements.

The octagonal lens is also able to act as a multi-beam antenna with 8 directive beams. The crossover points of two adjacent radiation patterns are located at a 22.5$\degree$ offset with respect to the centre of each segment. At these points, the phase of the radiated fields from two adjacent feed points is equal. When adjacent ports are fed in simultaneous anti-phase the electric fields propagating in these directions completely cancel each other out. The resulting radiation pattern is a multi-beam profile with directive beams emerging from each of the 8 lens segments, as shown in \cref{Fig3a}. Each beam has a half-power beamwidth of 20$\degree$ and are separated by zero-field nulls.
When all 8 elements are excited with equal phase and amplitude, the directive beam pattern from \cref{Fig3a} can effectively be rotated by 22.5$\degree$ as the electric fields in the off-axis directions combine in-phase to form a maximum. \cref{Fig3b} shows the radiation patterns when each port is fed in equal phase and amplitude. The nulls separating adjacent beams are around 8 dB below the maximum level of the main beam.

One of the drawbacks of phased antenna arrays is achieving the required phase and amplitude weightings required to each of the antenna elements, due to complexity, cost, power consumption and system losses. Here however, the different radiation pattern combinations presented can be achieved through simple switched feeding networks. The individual sector patterns require only a single port to be activated at once which can easily be achieved with an 8-1 RF switch. The multi-beam patterns require all elements to be fed simultaneously, with either equal phase or a 180$\degree$ phase shift to adjacent ports, which can be achieved through common RF techniques such as delay-line phase shifters or rat-race couplers. 

\section{3D Printed Graded-Dielectric Infill Patterns}\label{sec4}
3D printing can be used to fabricate graded dielectric materials due to the ability to control the fill-factor ratio of filament to air-gap intrusions. A number of methods exist for estimating the effective dielectric constant of such a medium, such as the Maxwell-Garnett relationship, the arithmetic mean, and the harmonic mean. The realised values are however dependent on the size and structure of the unit-cell, which can be designed to achieve specific isotropic or anisotropic permittivity values \cite{doi:10.1002/admt.201600072,10.1038/ncomms1126,LARIMORE201748}. Any 3D printed structure can be printed with a specific fill-factor using default slicing patterns such as rectilinear grids or a hexagonal honeycomb lattice, which also allows effective permittivity values to be generated. The advantage of using default slicing patterns to vary the infill-ratio of 3D printed parts lie in the fact that models are far easier to design and fabricate when compared to customised unit cell patterns. 

\begin{figure}[!t]
	\centering
    \includegraphics[width=0.9\linewidth]{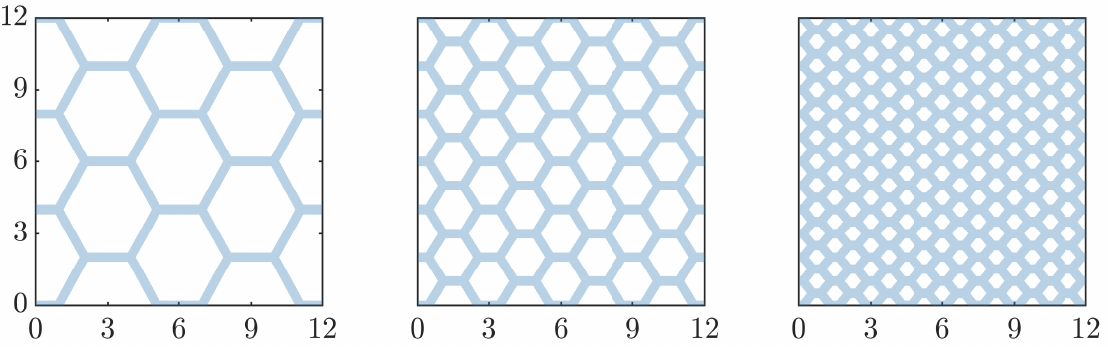}%
	\caption{Different filling factors of the hexagonal unit cell pattern auto-generated from slicing software. From left to right the filling factor ratio of filament (blue) to air (white) is 20\%, 37\% and 66\%.}
    \label{Fig4}%
	\vspace{0.5cm}
	\centering
	\raisebox{-0.5\height}{\includegraphics[width=0.9\linewidth]{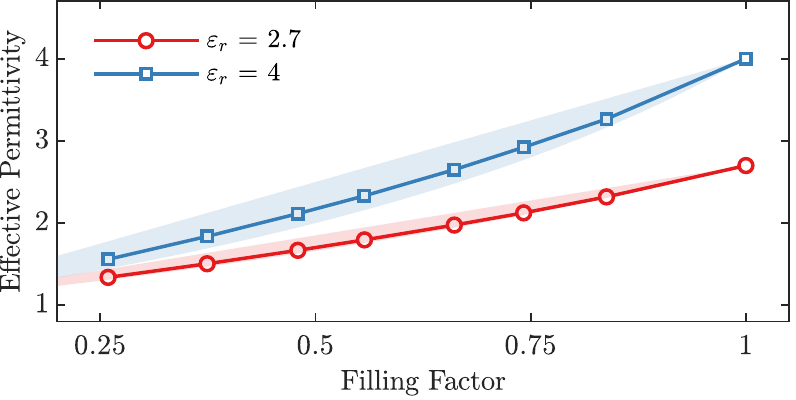}}%
	\caption{The effective permittivity of the hexagonal unit cell pattern at 5.8 GHz with varying filling factor and two different host permittivities. The shaded region shows the bounds calculated by the arithmetic mean and Maxwell-Garnett relationship}%
	\label{Fig5}%
\end{figure}

In fused deposition modeling (FDM) 3D printer technology, filament is extruded through a heated nozzle with a fixed diameter. The extrusion width of the filament is therefore defined by the size of the nozzle. The unit-cell size of automatically generated infill patterns increase as the fill-factor ratio of filament to air is reduced (i.e. the air-gap intrusions become larger). A hexagonal honeycomb pattern was chosen as the basis for controlling the fill-factor ratio of the 3D printed lens. \cref{Fig4} shows the honeycomb infill pattern with 3 different filling factors. The filament is represented by the blue lines which have a thickness of 0.4 mm. The first pattern has filling factor of 20\%, for the centre pattern it is 37\%, and the final pattern filling factor is 66\%. 

The relationship between the unit cell pattern and fill factor and the effective dielectric constant was determined through full-wave simulations. The hexagonal pattern was generated computationally for a number of different filling factor ratios between 0 and 100\%. The pattern was extruded in the $z$ direction and the effective permittivity was extracted using the s-parameter retrieval method \cite{PhysRevE.70.016608}, with Floquet ports and unit cell boundary conditions terminating the simulation space. Prior to the modelling, the dielectric permittivity of two different filaments, a standard white ABS from Verbatim and ABS-400 filament from Preperm \cite{PREPERM}, were characterised using a Keysight 85072A 10-GHz Split Cylinder Resonator. The standard ABS had a permittivity of 2.69 and ABS-400 had a permittivity of 3.99, with measured loss tangents of 0.002 and 0.0015 respectively. 
The extracted effective permittivity of different filling ratios for the two filaments are shown in \cref{Fig5}. As can be seen, the effective permittivity lies within the boundaries of the Maxwell-Garnett and arithmetic mean. The lowest achievable permittivity from the ABS-400 filament was 1.6 when the filling factor ratio of plastic to air was 0.25, whilst for the standard ABS it was equivalent to 1.3. In order to achieve the full range of permittivity values required for the discretized octagonal lens (ranging from 1.3 to 3.6), both filaments are required. The fill factor and host filaments required for achieving the effective permittivities of the discretized lens (\cref{Fig1c}) are displayed in \cref{Table1}.

\begin{table}[!t]
	\renewcommand{\arraystretch}{1.3}
	\caption{The required fill factor for achieving different effective permittivities of the octagonal lens.}
	\label{Table1}
	\centering
	\begin{center}
		\begin{tabular}{|C{2cm}|C{2cm}|C{2cm}|}
			\hline
			Effective $\varepsilon_r$ & Filament & Fill Factor\\
			\hline
			3.6 & ABS-400 & 87\%\\
			3.1 & ABS-400 & 75\%\\
			2.7 & ABS & 100\%\\
			2.3 & ABS & 80\%\\
			1.9 & ABS & 60\%\\
			1.6 & ABS & 35\%\\
			1.3 & ABS & 25\%\\
			\hline
		\end{tabular}
	\end{center}
\end{table}

\section{Fabrication, Modelling and Characterisation of 3D Printed Octagonal Multi-Beam Lens}\label{sec5}

\begin{figure}[!t]
	\centering
	\subfloat[]{%
		\includegraphics[height=3.5cm]{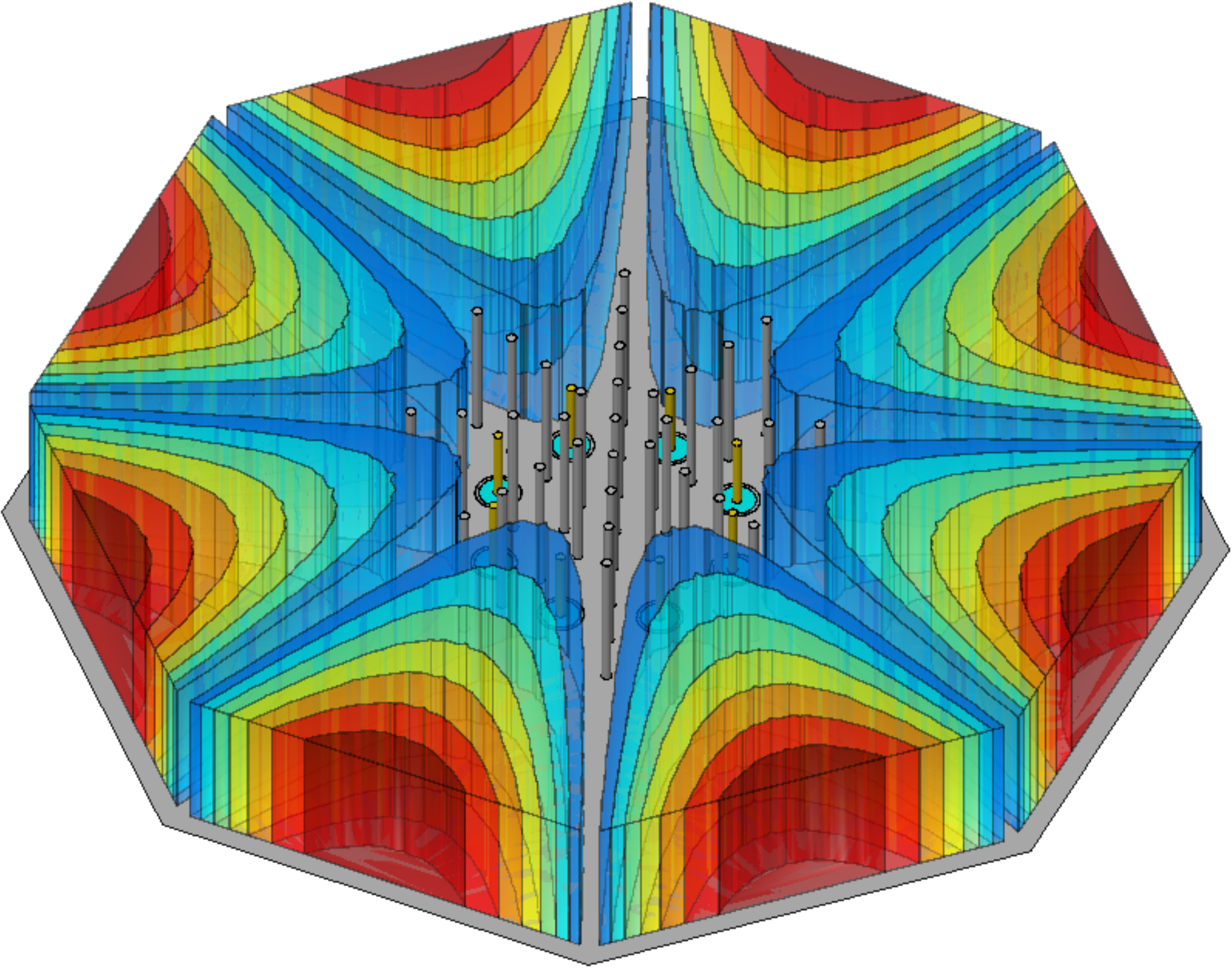}%
		\label{Fig6a}%
	}
	\hfill
	\subfloat[]{%
		\includegraphics[height=3.7cm]{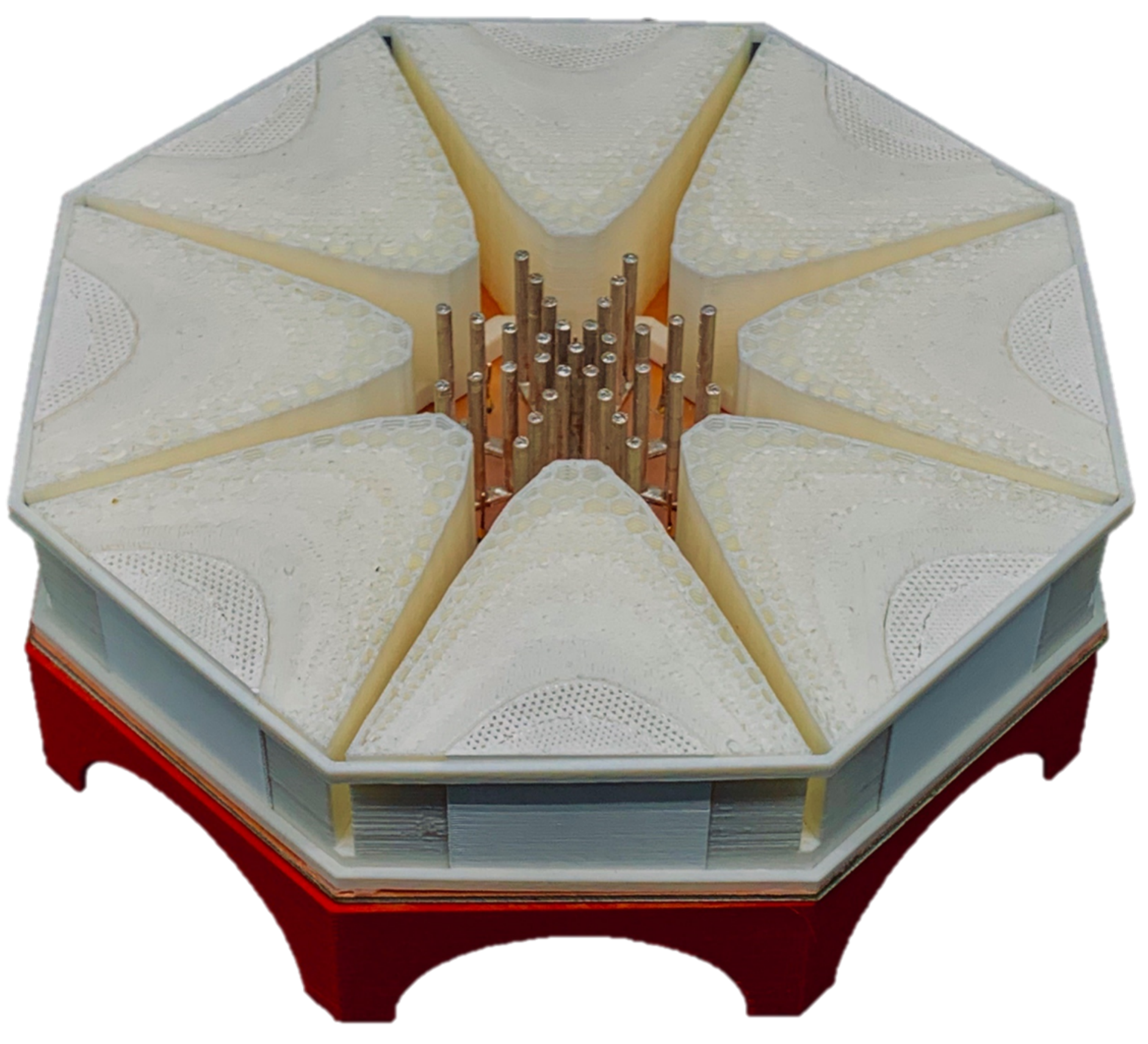}%
		\label{Fig6b}%
	}
	\caption{Full 3D antenna strucutre. (a) CAD model of 3D lens. (b) Photograph of 3D printed lens prototype with embedded feeding structure positioned on a metallic ground plane.}
	\label{Fig6}
\end{figure}

In order to verify that the proposed lens was suitable for realistic antenna applications, a full 3D antenna model was built and simulated. The permittivity map and reflector elements were extruded along the $z$ direction to a height of 25 mm and placed on an octagonal PEC ground plane with a width and length of 180 mm. At each of the feed points, a $\sfrac{\lambda}{4}$ monopole was positioned which were fed from 50 $\Omega$ coaxial lines beneath the ground plane. All other dimensions and material values were the same as the 2D simulations. The full 3D model can be seen in \cref{Fig6a}

For fabrication, the model of the 3D lens was split into different parts based on the effective permittivity profile, and two separate printing jobs were then created based on the host filament requirements of the different parts. The inner section of the lens, which was printed with a standard ABS filament, contained 6 different sectors with fill factor ranging from 100\% to 25\% covering permittivity values of 2.7 down to 1.3. Using the open source slicing software, Slic3r (v1.30), the model was prepared with a the appropriate infill density settings for each of the model parts. The same steps were repeated for the second part of the model requiring the ABS400 filament. The GRIN lens was then printed using an Original Prusa i3 MK3 3D printer with 0.4 mm diameter nozzle. The printed structure is shown in \cref{Fig6b}. SMA connectors were soldered to the rear side of an octagonal copper ground plane in the feed point positions. Strips of wire with a length of 13 mm were attached to the connectors to form the monopole feeds. The reflecting elements with diameter of 3 mm were 3D printed using the standard ABS filament and coated with silver plated copper shielding spray coat (sheet resistance, $R_s$ = 0.77 $\Omega/sq$), then hand painted with Electrolube silver conductive paint ($R_s$ = 0.03 $\Omega/sq$) to ensure high conductivity throughout.

\begin{figure}[!t]
	\centering
	\raisebox{-0.5\height}{\includegraphics[width=0.8\linewidth]{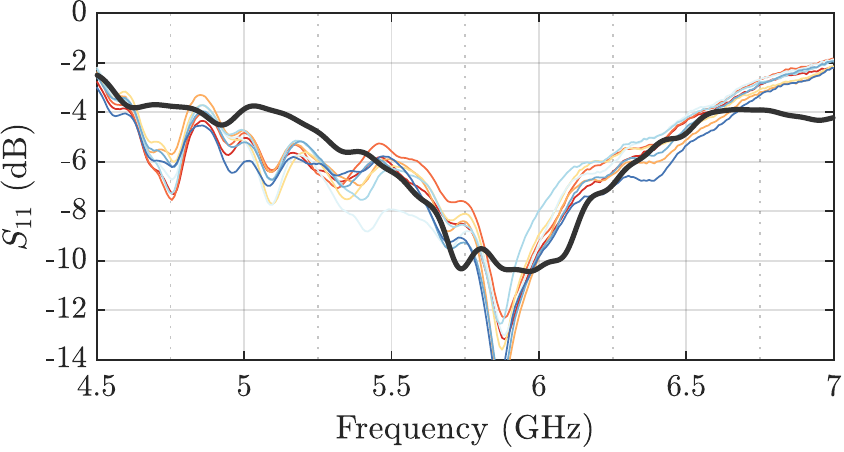}}%
	\caption{Input response of each of the feeding monopoles of the octagonal GRIN lens. The dark grey line shows the simulated data and coloured lines represent measured data of each port.}
	\label{Fig7}

	\captionsetup[subfigure]{labelformat=empty}
	\centering
	\hspace{0.5cm}
	\subfloat[E-plane]{%
		\includegraphics[height=3.5cm]{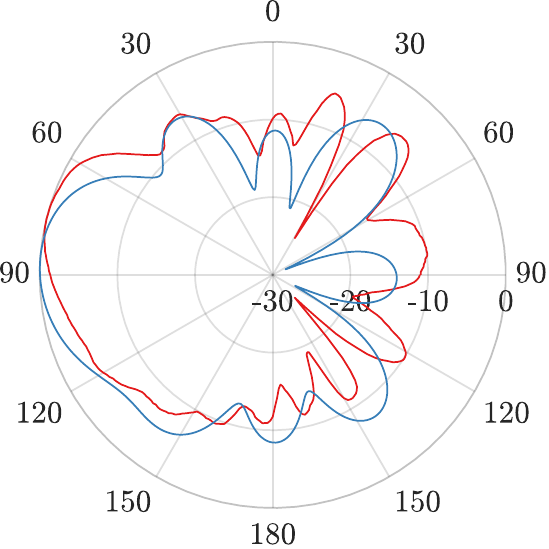}%
	}
	\hfill
	\subfloat[H-plane]{%
		\includegraphics[height=3.5cm]{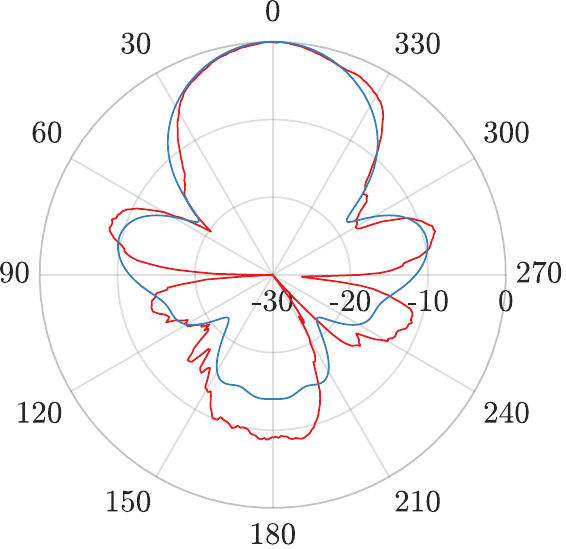}%
	}
	\hspace{0.5cm}
	\\
	\vspace{0.2cm}
	{\includegraphics[]{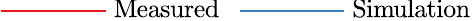}}%
	\caption{Measured and simulated E and H plane radiation patterns of a single sector of the octagonal GRIN lens at 5.8 GHz.}
	\label{Fig8}

	\captionsetup[subfigure]{labelformat=empty}
	\centering
	\hspace{0.5cm}
	\subfloat[Anti-phase]{%
		\includegraphics[height=3.5cm]{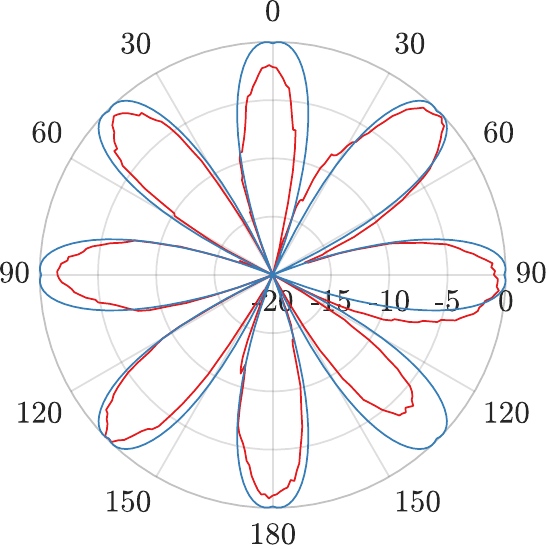}%
	}
	\hfill
	\subfloat[Equal-phase]{%
		\includegraphics[height=3.5cm]{Fig9-B.pdf}%
	}
	\hspace{0.5cm}
	\\
	\vspace{0.2cm}
	{\includegraphics[]{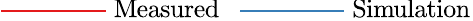}}%
	\caption{Measured and simulated H plane radiation patterns of them multi-beam octagonal GRIN lens at 5.8 GHz with anti-phase and equal phase excitation.}
	\label{Fig9}
	
\end{figure}

The simulated and measured input response of each of the 8 feed elements lens is shown in \cref{Fig7}. The input elements are matched to a return loss of below -10 dB between 5.8 and 6 GHz, and agree well with the simulated results. The mutual coupling between ports was strongest for adjacent ports, at a value of -16 dB at the centre frequency of 5.9 GHz, demonstrating a good isolation between feeding points due to the the reflector elements. The H and E plane radiation patterns of a single sector are displayed in \cref{Fig8}. The measured H-plane patterns have an -3 dB beamwidth of +/-23$\degree$ and fall to a -20 dB null at an angle of +/- 55$\degree$. Side lobes appear at +/-70$\degree$ with a strength 10 dB below the maximum. The back lobe has a signal strength which is 10 dB below the maximum. In the E-plane, the beam is tilted slightly upwards of the horizon, and the power level is above -3 dB for angles of 100$\degree$ $\textless$ $\theta$ $\textless$ 55$\degree$. The cross-polar patterns are not shown in \cref{Fig8}, however the maximum measured cross polar power level was 15 dB below the co-polar maximum, confirming that a strong vertically polarised field is radiated by the monopole feeding elements. As can be seen, the measured data agrees well with the simulations. The maximum gain of the antenna for single segment excitation, which was measured by comparing to a standard gain horn antenna, was 8.5 dBi.

The H-plane multi-beam radiation patterns were measured using a 24-port Rohde and Schwarz ZNBT-8 vector network analyser. The transmission response between a polarised feeding horn antenna and each of the 8 ports of the lens were measured simultaneously as the antenna under test was rotated through 360$\degree$. The resulting fields were summed together with anti and equal phase applied to adjacent ports. The measured and simulated patterns from the 3D model are displayed in \cref{Fig9}. In the measured data, there is a slight variation in the maximum normalised power of each beam by up to 3 dB which is due to the misalignment of the lens above the axis of rotation in the anechoic chamber. The maximum measured gain of the multi-beam antenna for anti-phase excitation was 4.9 dBi, and 5.9 dBi for equal-phase excitation.

\section{Operating as an Omnidirectional Radiator}\label{sec6}
The previous analysis of the proposed antenna focussed on its radiation pattern performance when the 8 element feed array was excited either individually or with phase weighting profiles that could be realised through simple feeding networks. However, modern radio terminals have the ability to excite large antenna arrays with phase and amplitude weighting to each individual port, typically for beam-steerable arrays. Here, the proposed antenna was designed to operate with a switched directional pattern which can be used in MANET radios where a mobile terminal may need to isolate a target in a particular direction. However such an antenna may also be required to radiate omnidirectionally at a given moment in order to determine the direction of the required targets. By applying the appropriate phase and amplitude weightings to each port, it is possible to generate an omnidirectional pattern from the octagonal lens.

Using both the simulated and measured H-plane patterns, the amplitude and phase weighting for each of the 8-ports in order for the antenna to produce an omnidirectional radiation pattern were determined through particle swarm optimization. The objective function was defined as 

\begin{equation}\label{eq:2}
obj = \min\Big(\int_{0}^{2\pi}|E_{total}(\phi) - E_{target}|d\phi\Big)
\end{equation}

\noindent where $E_{total}$ is the total radiation pattern obtained from summing the radiation patterns of each individual element, $E_n$ with amplitude and phase weightings $A_n$ and $\beta_n$ respectively. 
\begin{equation}\label{eq:3}
E_{total} = A_1e^{i\beta_1}E_1(\phi) ... + ... A_8e^{i\beta_8}E_8(\phi)
\end{equation}
\noindent In order to minimize the H-plane ripple, the value of $E_{total}$ used in the objective function was modified according to \cref{eq:4}, where $E_{target}$ = 0.7071. The swarm size was set to 50 and the optimization was terminated after 50 stall iterations.
\begin{equation}\label{eq:4}
E_{total}(\phi)(E_{total}(\phi) \textgreater E_{target}) = E_{target}
\end{equation}

\begin{figure}[!t]
	\centering
	\hspace{0.1cm}
	\subfloat[]{%
		\includegraphics[width=0.45\linewidth,valign=c]{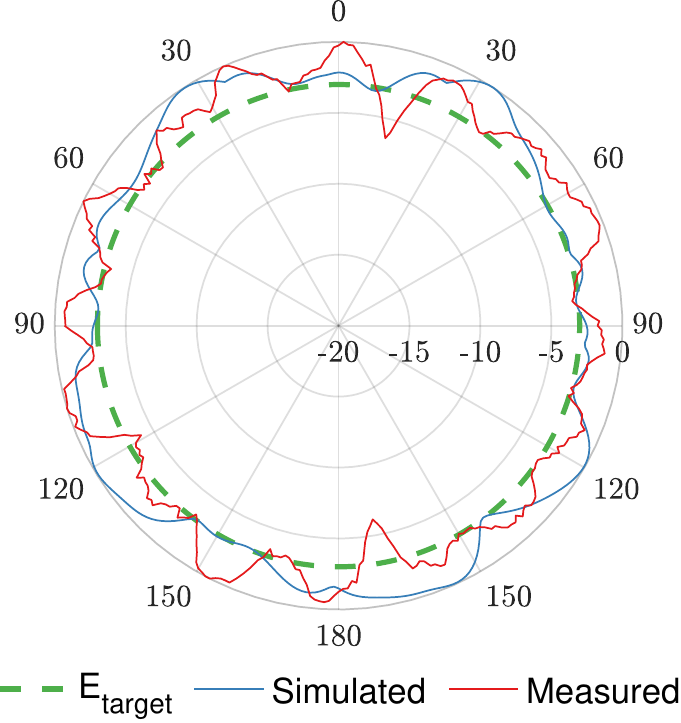}%
		\label{Fig10a}
	}
	\hfill
	\subfloat[]{%
		\includegraphics[height=3.1cm,valign=c]{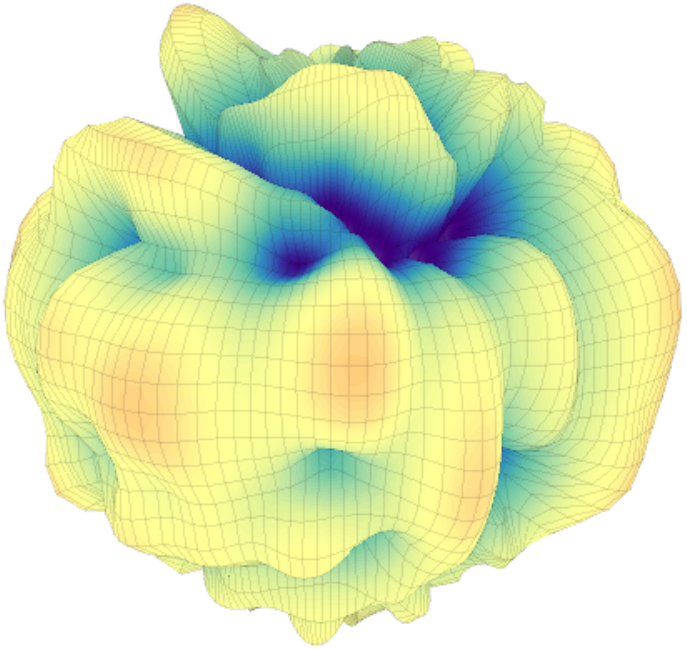}%
		\vphantom{\includegraphics[width=0.45\linewidth,valign=c]{Fig10-A.pdf}}%
		\label{Fig10b}
	}
	\includegraphics[height=3.4cm,valign=c]{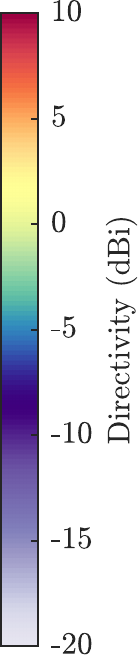}%
	\hfill
	\caption{Omnidirectional radiation patterns. (a) H-plane patterns with all power levels of above -3 dB calculated using \cref{eq:2,eq:3,eq:4}. (b) The 3D pattern from full-wave simulations using port the weightings of \cref{Table2}.}
	\label{Fig10}
	
\end{figure}

\begin{table}[!t]
	\renewcommand{\arraystretch}{1.3}
	\caption{Amplitude and phase weighting to produce omnidirectional radiation pattern - simulations (S) and measurements (M)}
	\label{Table2}
	\centering
	\begin{center}
		\begin{tabular}{|cc|cccccccc|}
			\hline
			\multicolumn{2}{|r|}{Port}& 1 & 2 & 3 & 4 & 5 & 6 & 7 & 8 \\
			\hline
			\hline
			\multirow{2}{*}{S}&$A$\ & 1.00 & 0.49 & 0.24 & 0.59 & 0.91 & 0.55 & 0.09 & 0.46 \\
			&$\beta$ & 0 & -127 & -94 & -6 & -73 & 1 & -155 & -70 \\
			\hline
			\hline
			\multirow{2}{*}{M}&$A$ & 1.00 & 0.23 & 0.87 & 0.25 & 0.99 & 0.34 & 0.99 & 0.13 \\
			&$\beta$ & 0 & -110 & -14 & 225 & 46 & 178 & -16 & -110 \\
			\hline
		\end{tabular}
	\end{center}
\end{table}

\cref{Fig10a} shows the H-plane radiation patterns of $E_{target}$ and the optimised $E_{total}$ result computed using \cref{eq:2,eq:3,eq:4}. The H-plane pattern of $E_{total}$ is omnidirectional with all power levels above the -3 dB limit set by $E_{target}$, i.e. there is a less than 50\% variation in the magnitude of the radiated power level. The optimised amplitude and phase weightings are shown in \cref{Table2}. In order to confirm the omnidirectional performance of the radiation pattern, the $A_n$ and $\beta_n$ values from \cref{Table2} were applied to the ports of the full-wave 3D antenna model with simultaneous excitation. \cref{Fig10b} shows the 3D patterns which is generally omnidirectional and has a maximum directivity of 3.9 dBi.
Finally, the omnidirectional H-plane pattern was also calculated from the measured patterns of each port as previously described. The particle swarm optimisation was re-run using the measured data after each pattern was normalised to its maximum power level. The resulting omnidirectional pattern is shown in \cref{Fig10a} where it can be seen that a similar omnidirectional performance is obtained to the simulated patterns. The resulting amplitude and phase weightings applied to the measured patterns are listed in \cref{Table2}.

\section{Conclusions}\label{sec7}
In summary, this paper has presented the application of a  graded-dielectric lens antenna fed by a switched monopole feed array to achieve pattern diversity over 360$\degree$ in azimuth at 5.8 GHz. It was shown how the lens can radiate with multiple equally spaced beams when excited with the appropriate phase. The lens was also able to radiate with an omnidirectional pattern, despite the directive nature of each individual sector. The lens was fabricated using multi-material 3D printing and the effective medium theory to achieve a graded dielectric profile from 3.6 down to 1.3. The proposed antenna would be suitable for application in a MANET radio to be mounted on a mobile terminal when directive beams are required for interference mitigation and targeted communications. 
\bibliographystyle{IEEEtran}
\bibliography{8-beam-lens_Bibtex}%
\end{document}